\documentclass [11pt]{article}
\usepackage{amsmath,amssymb,epsfig,cite}

\numberwithin{equation}{section}

\begin{document}

\begin{flushright}
IISc/CHEP/10/07\\
\end{flushright}
\begin{center}
{\bf THE QUANTUM SINH-GORDON MODEL IN NONCOMMUTATIVE (1+1) DIMENSIONS}
\bigskip

Sachindeo Vaidya\footnote{vaidya@cts.iisc.ernet.in} \\
\smallskip
{\it Centre for High Energy Physics, \\
Indian Institute of Science, Bangalore 560 012, INDIA.}
\end{center}

\begin{abstract}
Using twisted commutation relations we show that the quantum
sinh-Gordon model on noncommutative space is integrable, and compute
the exact two-particle scattering matrix. The model possesses a
strong-weak duality, just like its commutative counterpart.
\end{abstract}

\section{Introduction}
The study of quantum field theory (QFT) in $(1+1)$-dimensions provides
valuable insights into many difficult conceptual problems of QFT's, as
well as giving us situations where appropriate use of infinite
dimensional symmetries, like conformal invariance and $W_\infty$
symmetry, considerably simplifies the field theoretic computation. One
might ask if similar insights could also be obtained by studying QFT's
in $(1+1)$-dimensional noncommutative spacetime. Indeed there is
reason for some optimism regarding this query, as it is possible to
define notions of conformal invariance, Kac-Moody and Virasoro
symmetries \cite{lvv,bmqt}.

In this article we will address the question of quantum integrability,
and investigate the noncommutative analog of the quantum sinh-Gordon
model. We will argue that it is integrable, in the sense that there is
no particle production, and calculate the exact two-particle scattering
matrix.

In Section 2, we will summarize the results on twisted QFT's based on
our earlier work \cite{bmpv,replyto}, and then review scattering
theory for noncommutative QFT's \cite{bpq}. In Section 3, we will
recall some essential features of the quantum sinh-Gordon model on
commutative spacetime, and then go on to construct the two-particle
$S$-matrix $S_\Theta$ for its noncommutative counterpart. A discussion
of the properties of $S_\Theta$ will be provided in Section 4.

\section{Noncommutative $(1+1)$-d spacetime}
Two dimensional noncommutative spacetime is generated by operators
$\hat{x}_\mu$ satisfying the commutation relations
\begin{equation} 
[\hat{x}_\mu, \hat{x}_\nu] = i \Theta_{\mu\nu} \equiv i \Theta
\epsilon_{\mu\nu}, \quad \mu,\nu = 0,1 \ , \label{ncrelation}
\end{equation} 
where $\hat{x}_0, \hat{x}_1$ are hermitian operators, and $\Theta$ is
a real constant.

Using the Moyal map, we can map operators built out of (the products
of) the $\hat{x}_\mu$'s to functions on Minkowski space ${\mathbb
R}^{1,1}$, but with a modified rule for multiplication. If $\hat{f}$
and $\hat{g}$ are two operators that are mapped to functions $f(x)$
and $g(x)$ respectively, then their {\it star-product} consistent with
(\ref{ncrelation}) is
\begin{equation} 
(f*g)(x) = f(x) e^{\frac{i}{2} \overleftarrow{\partial}^\mu
    \Theta_{\mu\nu} \overrightarrow{\partial}^\nu} g(x) \ .
\end{equation} 
For example, for $e_p(x) = e^{-i p \cdot x}$,
\begin{equation} 
e_p(x) * e_q(x) = e^{-\frac{i}{2}p \wedge q} e_{p+q}(x), \quad p
\wedge q \equiv p^\mu \Theta_{\mu\nu} q^\nu \ .
\label{planewavestar}
\end{equation}   
Notice that the commutation relation (\ref{ncrelation}) is unchanged under
translations $\hat{x}_\mu \rightarrow \hat{x}_\mu + a_\mu$ and
identity-connected Lorentz transformations $\hat{x}_\mu \rightarrow
\Lambda_\mu^{\phantom{\nu}\nu} \hat{x}_\nu$. Thus Poincar\'e
transformations, with their usual action on coordinates, belong to the
automorphism group of noncommutative ${\mathbb R}^{1,1}$. 

Our interest in this noncommutative space is to understand the notion
of integrability for quantum theories. As long as the Hamiltonian is
(formally) hermitian, the quantum theory on such a space is manifestly
unitary, as emphasized in the case of field theories by \cite{bdfp},
and demonstrated for single particle quantum mechanics by
\cite{bgmt}. We will therefore proceed to write free quantum fields,
which we will use to build interactions.

\subsection{Quantum Fields}

A free quantum scalar field $\Phi(x)$ can be expanded as
\begin{equation} 
\Phi(x) = \int d \mu (k) (a_k e^{-i k \cdot x} + a^\dagger_k e^{i k
  \cdot x}), \quad d\mu(k) = \frac{dk^1}{4 \pi k^0}, \quad k^0 =
  \sqrt{(k^1)^2 + M^2} \ .
\label{tfield}
\end{equation} 
While it is possible to quantize the field by imposing the standard
(or canonical) commutation relations between $a_k$'s and
$a^\dagger_k$'s, more general commutation relations are possible as
well \cite{bmpv}. These {\it twisted} commutation relations are of the
form:
\begin{eqnarray}  
a_p a_q &=& G_{p,q} a_q a_p, \label{twist1}\\
a_p a^\dagger_q &=& G_{p,-q}a^\dagger_q a_p + 2p_0 \delta (\vec{p}
- \vec{q}), \label{twist2}\\
a^\dagger_p a^\dagger_q &=& G_{-p,-q}a^\dagger_q a^\dagger_p
\label{twist3} 
\end{eqnarray} 
where $G_{p,q}$ is any Lorentz-invariant function of two-momenta $p$
and $q$.

In \cite{replyto}, we argued that in spacetime dimension greater than
2, compatibility with quantum statistics of identical particles
requires $G_{p,q}$ to be of the form
\begin{equation} 
G_{p,q} = e^{i p \wedge q} \ . 
\label{Gform}
\end{equation} 
We will work henceforth with this choice of $G_{p,q}$, and argue in
Section 3 why this leads to integrability of the $S$-matrix for the
noncommutative sinh-Gordon model.

Even though the commutation relations
(\ref{twist1}--\ref{twist3}) are twisted, the operators
$a_p, a^\dagger_q$ act on the usual (bosonic) Fock space. To see this,
let $c^\dagger_p, c_q$ be the ordinary or untwisted creation and
annihilation operators:
\begin{equation} 
[c_p,c_q] = 0, \quad [c_p,c^\dagger_q] = 2p_0 \delta(\vec{p} -
\vec{q}). \label{usualcomm}
\end{equation} 
There is a simple relation between the $c_p$'s and the $a_p$'s (for
$G_{p,q} = e^{i p \wedge q}$). To this end, consider the Fock space
momentum operator $P_\mu$ associated with $\Phi(x)$:
\begin{eqnarray} 
P_\mu &=& \int d \mu(p) p_\mu a^\dagger_p a_p \ , \\
{[}P_\mu, \Phi(x)] &=& -i\partial_\mu \Phi(x)
\end{eqnarray}  
Using $P_\mu$, we can realize the twisted creation-annihilation
operators in terms of the untwisted ones:
\begin{eqnarray} 
a_p &=& c_p e^{-\frac{i}{2}p \wedge P}, \label{atoc1}\\
a^\dagger_p &=& c^\dagger_p e^{\frac{i}{2}p \wedge P} \label{atoc2},
\end{eqnarray} 
It is easy to check that (\ref{atoc1}, \ref{atoc2}) reproduce the
twisted commutation relations (\ref{twist1}--\ref{twist3}). Thus the
$a_p$'s act on the same Fock space as that of the untwisted creation
and annihilation operators. The map (\ref{atoc1},\ref{atoc2}) is a
``dressing transformation'' (first discussed in \cite{grosse}), and
the commutation relations for the $a_p$'s and $a^\dagger_p$'s that
follow from it are simple examples of the algebra discussed
in\cite{zz1,faddeev}.

A twisted $n$-particle state is 
\begin{equation} 
|p_1,p_2, \cdots p_n \rangle_\Theta = a^\dagger_{p_1} \cdots
 a^\dagger_{p_n} |0\rangle
\end{equation}  
This is related very simply to the ordinary $n$-particle state. Using
(\ref{atoc2}), 
\begin{equation} 
|p_1,p_2, \cdots p_n \rangle_\Theta = e^{\frac{i}{2} \sum_{i<j} p_i
 \wedge p_j} |p_1,p_2, \cdots p_n \rangle_0 \label{statemap}
\end{equation}
In $(1+1)$ dimensions, it is often convenient to work in light-cone
coordinates, with the two-momentum characterized by {\it rapidity}
$\eta$: $(p^0,p^1) = (M \cosh \eta, M \sinh \eta)$. The twisted
commutation relation (\ref{twist3}) then becomes
\begin{equation} 
a^\dagger (\eta_1) a^\dagger (\eta_2) = e^{-i \Theta M^2 \sinh (\eta_1
  - \eta_2)} a^\dagger (\eta_2) a^\dagger (\eta_1)
\end{equation} 

Finally, the quantum field $\Phi(x)$ even though free, is not local:
the commutator $[\Phi(x),\Phi(y)]$ is non-zero for $x$ and $y$
space-like separated \cite{replyto}.

\subsection{Noncommutative Scattering Theory}

Consider a theory with interaction Hamiltonian 
\begin{equation} 
H_I = g \int dx \Phi_*^n \ , \quad \Phi_*^n = \underbrace{\Phi(x) *
  \Phi(x) \cdots *\Phi(x)}_{n \,\,{\rm terms}} \ .
\end{equation} 
The scattering operator for this theory is 
\begin{equation} 
S_\Theta = T e^{-i\int H_I dt}
\end{equation} 
The first non-trivial term in the perturbative expansion of the above
is 
\begin{equation} 
S_\Theta^{(1)} = -i g \int d^2 x \Phi_*^n
\end{equation} 
Using the mode expansion for $\Phi(x)$, let us look at a typical term
in the above, which is of the form $-ig \int d^2 x a_{p_1} \cdots
a_{p_n} e_{p_1}(x)* \cdots e_{p_n}(x)$. Using (\ref{atoc2}) and
(\ref{planewavestar}),
\begin{eqnarray}  
&&-ig \int d^2 x a_{p_1} \cdots a_{p_n} e_{p_1}(x)* \cdots e_{p_n}(x) \\ 
&=& -ig \int d^2 x c_{p_1} \cdots c_{p_n} e^{-\frac{i}{2} (\sum_i p_i)
   \wedge P} e^{\frac{i}{2} \sum_{i<j} p_i \wedge p_j} e_{p_1 +p_2
     +\cdots p_n}(x) e^{-\frac{i}{2} \sum_{i<j} p_i \wedge p_j} \\
&=& -ig \int d^2 x c_{p_1} \cdots c_{p_n}e_{p_1 +p_2 +\cdots p_n}(x)
  e^{\frac{1}{2}\overleftarrow{\partial} \wedge P} \\
&=& -ig \int d^2 x c_{p_1} \cdots c_{p_n} e_{p_1}(x) \cdots e_{p_n}(x)
  \label{cS} 
\end{eqnarray}
where we have integrated by parts and discarded surface terms to
obtain the last expression. But (\ref{cS}) is just the same as the
corresponding term from the commutative scattering theory. Hence to
order $g$, the noncommutative scattering operator is the same as in
the commutative theory.

More generally, this is true to any order in perturbation theory (see
\cite{bpq} for details of the proof). In particular, this means that
\begin{equation} 
S_\Theta = S_0 \ .
\label{Soperator}
\end{equation} 
The scattering operator on noncommutative space is the same as that
for the commutative counterpart. In particular, if a commutative
theory is renormalizable, so is its noncommutative counterpart,
because the number of counterterms is the same. 

As far as scattering is concerned, the main difference between a
commutative theory and its noncommutative counterpart is in the nature
of asymptotic states. But since we know the explicit map
(\ref{statemap}) between these two kinds of asymptotic states, the
matrix elements of $S_\Theta$ can be calculated in terms of those of
$S_0$.

The result (\ref{Soperator}) is true for theories without interacting
non-Abelian gauge fields only \cite{bpqv}. For example, in
noncommutative $QCD$ there are effects that violate Lorentz
invariance. However, this caveat is not applicable here, as we will
only consider theories where the interaction Hamiltonian is made up of
matter fields only.

\section{Noncommutative sinh-Gordon Model}
One of the the simplest non-trivial integrable model in
$(1+1)$-dimensional commutative spacetime is the sinh-Gordon
model. This is the theory with interaction Hamiltonian of the form
\begin{eqnarray} 
H_I^{(sG)} &=& {\bf :}\int dx \left(\frac{M^2}{2} \Phi^2 +
\frac{1}{4!} \Phi^4 + \frac{g^2}{6!} \Phi^6 + \cdots \right) {\bf :} \\
&=& {\bf :}\int dx \frac{M^2}{g^2} \left(\cosh g \Phi -1 \right) {\bf :}
\end{eqnarray} 
where the double dots above stand for normal-ordering. For this
theory, there is no particle production: the amplitude $S_{m
\rightarrow n}$ for producing $n$ outgoing particles by colliding $m$
incoming particles in zero if $m \neq n$. In addition, the ``elastic''
amplitude $S_{m \rightarrow m}$ factorizes into products of
two-particle scattering amplitudes $S_{2\rightarrow 2}$. The exact
expression of the two-particle $S$-matrix on commutative spacetime has
been given in \cite{afz}:
\begin{equation} 
S_0 (\eta) = \frac{\tanh \left[\frac{1}{2}\left(\eta -i
    \frac{\pi}{2}B(g)\right) \right]}{\tanh \left[\frac{1}{2} \left(\eta + i
    \frac{\pi}{2}B(g)\right) \right]}, \quad {\rm where} \quad B(g) =
    \frac{2g^2}{8\pi + g^2} 
\end{equation} 
and $\eta$ the relative rapidity. 

A very nice argument motivating quantum integrability for the
commutative case has been given by Dorey in \cite{dorey}. We will use
this argument, adapting it appropriately to the noncommutative case.
\begin{figure}
\centerline{\epsfig{figure=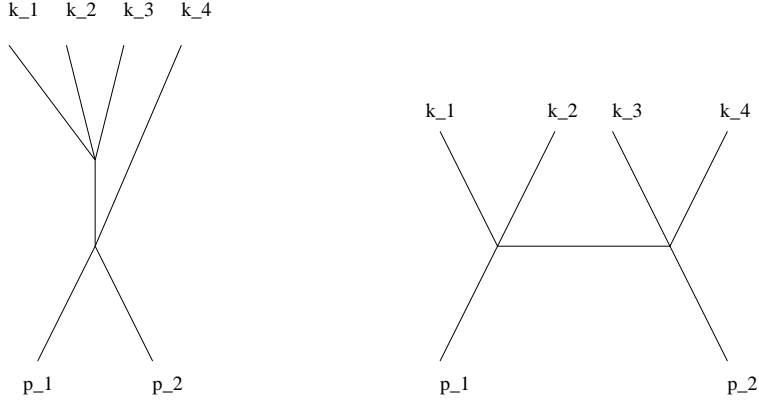,clip=9cm,width=10cm}}
\caption{Tree-level Feynman diagrams for $2\rightarrow 4$ process in
  $\Phi_*^4$ theory}
\label{fig:2to4-1}
\end{figure}

Consider the theory based on the free field $\Phi(x)$ as in
(\ref{tfield}), with the free Hamiltonian $H_0 = \int d\mu(k)
a^\dagger_k a_k$, and an interaction $H^{(1)}_{int} =
:\frac{\lambda}{4!}  \int dx \Phi^4_*(x) :$, where by $\Phi^4_*(x)$ we
mean $\Phi(x)*\Phi(x)*\Phi(x)*\Phi(x)$.

There are two diagrams (Figure \ref{fig:2to4-1}) that contribute to the
$2\rightarrow 4$ scattering amplitude at tree level. The amplitude for
this process is
\begin{eqnarray} 
S_\Theta (p_1,p_2;k_1,k_2,k_3,k_4) &=& \langle 0 | a_{k_1} a_{k_2}
  a_{k_3} a_{k_4} | (-i \lambda) \left( \int d^2 x H_I (x,t) \right)
  |a^\dagger_{p_1} a^\dagger_{p_2}|0 \rangle \\ 
&=& S_0 (p_1,p_2;k_1,k_2,k_3,k_4) e^{\frac{i}{2} p_1 \wedge p_2 -
  \sum_{i<j} k_i \wedge k_j} \\ 
&=& i \frac{\lambda^2}{M^2} e^{\frac{i}{2} p_1 \wedge p_2 - \sum_{i<j}
  k_i \wedge k_j}
\end{eqnarray} 
where we have used (\ref{statemap}, \ref{Soperator}), and the fact that
$S_0 = i \frac{\lambda^2}{M^2}$.

Let us add an extra interaction of the form $H^{(2)}_{int} =
\frac{\lambda'}{6!} \int dx \Phi^6_*$. 
\begin{figure}
\centerline{\epsfig{figure=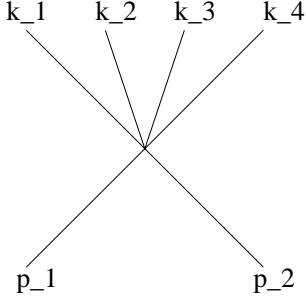,clip=3cm,width=4cm}}
\caption{Additional contribution to the $2\rightarrow 4$ process from the
  $\Phi_*^6$ term}
\label{fig:2to4-2}
\end{figure}
We see that the $2\rightarrow 4$ process receives an additional
contribution from a third diagram (Figure \ref{fig:2to4-2}), which of the
form
\begin{equation} 
S'_\Theta = -i \lambda' e^{\frac{i}{2} p_1 \wedge p_2 -
  \sum_{i<j} k_i \wedge k_j}
\end{equation} 
By choosing $\lambda' = \lambda^2/M^2$, we can make the total tree-level
amplitude $S_\Theta + S'_\Theta$ for the $2\rightarrow 4$ process to
vanish.

For the new interaction Hamiltonian $H_I^{(1)} + H_I^{(2)}$, the
amplitude for the $2\rightarrow 6$ process is now a non-zero constant,
which can be made to vanish by judiciously choosing an extra
interaction piece of the form $\int dx \Phi^8_*$. Continuing in this
manner, one finds that the theory with the interaction Hamiltonian of
the form
\begin{equation} 
H_I = {\bf :} \frac{M^2}{g^2} \int dx \left(\cosh_* (g \Phi) - 1
\right){\bf :} 
\label{ncsG}
\end{equation} 
has no particle production at tree-level: all processes of the form
$2\rightarrow n$ $(n>2)$ are forbidden (We have moved the mass term
$M^2\Phi^2/2$ from the free Hamiltonian to the interaction
Hamiltonian, so that it can be presented in the convenient form
(\ref{ncsG}).)

A crucial ingredient in the argument above is the specific choice
(\ref{Gform}) of the twist function $G_{p,q}$ for the commutation
relations of the free field creation and annihilation operators. Had
we chosen it to be of some other form, it is easy to see that we would
lose the {\it no particle production} condition, and hence quantum
integrability. In particular, choosing the twist function to be
identity (i.e. using conventional commutation relations
(\ref{usualcomm}) for the free field creation/annihilation operators)
leads to particle production at tree-level itself, as shown explicitly
by \cite{cm} (however, see also \cite{gp,gmpt,lmppt} for a discussion
of the absence of tree-level particle production in the noncommutative
{\it sine}-Gordon model).

Our argument also extends to higher loops. For a scalar field theory
(with, say, polynomial interactions) in commutative $(1+1)$
dimensions, the only source of ultraviolet divergence in perturbation
theory is from single closed loops (see for example
\cite{coleman}). These can be absorbed by renormalizing the mass of
the particle, or equivalently by working with the normal-ordering the
interaction Hamiltonian to start with, as we have done.

As we argued earlier, although the scattering operator for the
noncommutative theory and its commutative counterpart is the same, the
asymptotic states are different. Using the map (\ref{statemap}), we
find the two-particle amplitude for the noncommutative case to be
\begin{equation} 
S_\Theta (\eta) = \frac{\tanh \left[\frac{1}{2}\left(\eta -i
    \frac{\pi}{2}B(g)\right) \right]}{\tanh \left[\frac{1}{2}
    \left(\eta + i \frac{\pi}{2}B(g)\right) \right]} e^{-i\Theta M^2
    \sinh \eta}
\end{equation} 
where $M$ is the (physical) mass of the sinh-Gordon particle. 

\section{Discussion}

It is obvious that $S_\Theta(\eta)$ satisfies the following conditions:
\begin{description}
\item{Real analyticity:} $S_\Theta (\eta)$ is real for $\eta$ purely
  imaginary. 
\item{Unitarity:} $S_\Theta(\eta) S_\Theta(-\eta) = 1$.
\item{Crossing:} $S_\Theta(\eta) = S_\Theta(i\pi - \eta)$   
\item{Yang-Baxter equation:} $S_\Theta (\eta_{12})S_\Theta
  (\eta_{13})S_\Theta (\eta_{23}) = S_\Theta (\eta_{23})S_\Theta
  (\eta_{13})S_\Theta (\eta_{12})$ 
\end{description}
It also possesses the strong-weak duality symmetry: $S_\Theta$ is
invariant under $B \rightarrow 2-B$, or equivalently under
\begin{equation} 
g \rightarrow \frac{8\pi}{g} \ .
\end{equation} 

The fact the noncommutative scattering matrix for the sinh-Gordon
model differs from commutative one only by a phase may seem surprising
at first. It was pointed out by Mitra \cite{mitra} that overall phases
of the form $e^{i\sum_{\ell=0}^\infty b_{\ell} \sinh (2\ell+1) \eta}$ are
allowed, over and above the form of the $S$-matrix dictated by
dynamics. For local fields, the $b_{\ell}$ are required to vanish. However,
since our field is non-local, we have no such restriction. We find, in
fact, that $b_{\ell} = -\Theta M^2$.

\end{document}